\documentstyle[12pt]{article}
\makeatletter
\newcommand{\singlespacing}{\let\CS=\@currsize\renewcommand{\baselinestretch}
{1.0}\tiny\CS}
\newcommand{\doublespacing}{\let\CS=\@currsize\renewcommand{\baselinestretch}
{1.5}\tiny\CS}

\begin{document}
\title{Qubit rotation and Berry phase}
\author{Dipti Banerjee*\\
\singlespacing Rishi Bankim Chandra College, Department of
Physics,Naihati,\\24-Parganas(N), Pin-743 165,West Bengal,INDIA.\\ and \\
{\bf The Abdus Salam International Center for Theoretical Physics},\\
Trieste, Italy\\ and \\
Pratul Bandyopadhyay \\
Indian Statistical Institute,203 B.T.Road, Calcutta-700035, INDIA}
\date{}
\maketitle

\abstract{A quantized fermion can be represented by a scalar
particle encircling a magnetic flux line. It has the spinor
structure which can be constructed from quantum gates and qubits.
We have studied here the role of Berry phase in removing dynamical
phase during one qubit rotation of a quantized fermion. The
entanglement of two qubit inserting spin-echo to one of them
results the change of Berry phase that can be considered as a
measure of entanglement. Some effort is given to study the effect
of noise on the Berry phase of spinor and their entangled states.}
%\end{abstract}

\vspace{3cm}
 *Regular Associate of ICTP, Communicating author
email: deepbancu@hotmail.com.

\newpage

\section{\bf Introduction}

Entanglement is one of the basic aspects of quantum mechanics.It
was known long ago that quantum mechanics exhibits very peculiar
correlations between two physically distant parts of the total
system. Afterwards, the discovery of Bell's inequality (BI) [1]
showed that BI can be violated by quantum mechanics but have to be
satisfied by all local realistic theories. The violation of BI
demonstrates the presence of entanglement [2].

It is well known that the geometrical phase, such as the Berry
phase [3] play an important role in quantum mechanics. The
geometric origin of this phase is given by the holonomy of the
line bundle of the states where the phase emerges from the
integral of the connection (or curvature) of the bundle over the
parameter space [4].In recent years geometric phase in a single
particle system has been studied very well, both theoretically and
experimentally. The effect of Berry phase on entangled quantum
system is less known. But there is an interest to combine both
these quantum phenomenon[5].

It is natural to think that there is an inherent connection of
these two important quantum phenomena namely,entanglement and
Berry phase with quantization procedure. The aim of this paper is
to explore that connection.

Bell's inequality theorem may be interpreted as incompatibility of
requirement of locality with the statistical predictions of quantum
mechanics. So to study the Bell state, the role of {\it local}
spatial observations, apart from spin correlations, should also be
taken into considerations[6]. Since the Berry phase is acquired by
the space-time wave function when a particle traverses a closed path
and in the realm of quantum field theory it is associated with a
gauge field, the role of Berry phase becomes relevant in an
entangled  state. Indeed, chiral anomaly is a purely quantum effect
which arises due to the short distance singularity  and it is
expected that the influence of the Berry phase on an entangled state
is somehow linked up with that of the local observations of spins.
This suggests that to have a comprehensive view of the quantum
mechanical correlation of two spin 1/2 particles in an entangled
state we should take into account the role of the Berry phase
related to a spinor. In this note we shall study the formation of
spinor by the operation of quantum gates on qubits. We will
investigate the role of Berry phase in the one-qubit rotation in the
presence of the circulating magnetic field and  entangled state of
two qubit. Further we here also show some interest on the appearance
of noise in the topological phase of the pure and entangled state.

\section{\bf Quantization of a Fermi field and Berry Phase}

The quantization of a Fermi field can be achieved when we introduce
an anisotropy in the internal space through the introduction of a
direction vector as an internal variable[7]. The opposite
orientations of the direction vector correspond to particle and
antiparticle. To be equivalent to the Feynman path integral we have
to take into account complexified space-time when the coordinate is
given by $Z_\mu=x_\mu+i\xi_\mu$ where $\xi_\mu$ corresponds to the
direction vector attached to the space-time point $x_\mu$. Then the
field function will be of the form $\phi (x_\mu,\xi_\mu)$ which may
describe a particle moving in an anisotropic space. If $\chi$ be the
angle to specify the rotational orientation around the direction
vector $\xi_\mu$, then the wave function will depend on another
quantity $\mu$ apart from the coordinates $r, \theta,\phi$ where
$\mu$ is the eigenvalue of the operator $i\frac{\partial}{\partial
\chi}$. Here $\mu$ corresponds to the measure of anisotropy and
behaves like the strength of a magnetic monopole.Indeed in this
space the angular momentum is given by
\begin{equation}
\vec{J}=\vec{r}\times\vec{p}- \mu\hat{r}
\end{equation}

with $\mu=0,\pm1/2,\pm1...$ This corresponds to the motion of a
charged particle in the field of a magnetic monopole.

The spherical harmonics incorporating the term $\mu$ may be written
as [8]
\begin{equation}
{Y_l}^{m,
\mu}=(1+x)^{-(m-\mu)/2}.(1+x)^{-(m+\mu)/2}\times\frac{d^{l-m}}{d^{l-m}x}[(1+x)^{l-\mu}.
(1-x)^{l+\mu}]e^{im\phi}e^{i\mu\chi}
\end{equation}

with $x=cos\theta$.\\
 Since the chirality is associated with the
angle $\chi$ denoting the rotational orientation around the {\it
direction vector} $\xi_\mu$, the variation of the angle $\chi$ i,e
 will correspond to the change in chirality. In spherical harmonics given
by eqn.(2) the angular part associated with the angle $\chi$ is
given by $e^{-i\mu\chi}$. Thus when $\chi$ is changed to
$\chi+\delta\chi$, we have [9]
\begin{equation}
i\frac{\partial}{\partial(\chi+\delta\chi)}e^{-i\mu\chi}=
i\frac{\partial}{\partial(\chi+\delta\chi)}e^{-i\mu(\chi+\delta\chi)}e^{i\mu\delta\chi}
\end{equation}

which implies that the wave function will acquire the extra phase
$e^{i\mu\delta\chi}$ due to infinitesimal change of the angle
$\chi$ to  $\chi+\delta\chi$. When the angle $\chi$ is changed
over the closed path $0\leq\chi\leq2\pi$, for one complete
rotation, the wave function will acquire the phase
\begin{equation}
exp[ i\mu{{\int_0}^{2\pi}}\delta\chi] = e^{2i\pi\mu}
\end{equation}

which represents the Berry phase. Indeed in this formalism, a
fermion is depicted as a scalar particle moving in the field of a
magnetic monopole and $\mu=1/2$ corresponds to one flux quantum.
When a scalar field(particle) traverses a closed path with one flux
quantum enclosed, we have the phase $e^{i\pi}$ which suggests that
the system represents a fermion.

In general this Berry phase is the solid angle subtended by the
particle that can be seen by considering the two component spinor
structure of quantized spin $1/2$ particle. For the specific case of
$l=1/2,|m|=|\mu|=1/2$ from the spherical harmonics, we can construct
the instantaneous eigenstates $|\uparrow,t >$ as
\begin{equation}
|\uparrow,t> = {u\choose v} = {{Y_{1/2}}^{1/2,-1/2} \choose
{Y_{1/2}}^{-1/2,-1/2}} = {\cos\frac{\theta}{2}\exp i(\phi-\chi)/2
\choose \sin\frac{\theta}{2}\exp-i(\phi+\chi)/2}
\end{equation}
that can be written in terms of qubits $|0> and |1>$ as follows
\begin{equation}
|\uparrow,t> =[\cos\frac{\theta}{2} |0> +
\sin\frac{\theta}{2}e^{-i\phi} |1>]e^{i/2(\phi-\chi)}
\end{equation}

In similar to the coherent state approach [10], the effective
Lagrangian for this state becomes
\begin{equation}
L^\uparrow_{eff} = <\uparrow,t|\nabla_t|\uparrow,t> =
-\frac{i}{2}(\dot{\chi} - \dot{\phi}\cos\theta)
\end{equation}

The action integral over a closed path gives rise from the eqn.-7,
the required geometrical phase of the single quantized spinor.
\begin{equation}
\gamma_{\uparrow} = i \int L^\uparrow_{eff} dt = i\oint
A_{\uparrow}(\lambda)d\lambda= \frac{1}{2}(\oint
d\chi-\cos\theta\oint d\phi )= \pi(1 - \cos\theta)
\end{equation}
This shows that for quantized spinor the Berry Phase is a solid
angle subtended about the quantization axis. For $\theta=0$ the
minimum value of $\gamma_{\uparrow}$ is $0$ and at $\theta=\pi$
maximum.

The conjugate spinor of equation-6
\begin{equation}
|\downarrow(t)> = (\sin\frac{\theta}{2}|0> +
\cos\frac{\theta}{2}e^{i\phi}|1>)e^{-i/2(\phi-\chi)}
\end{equation}
possesses the Berry phase over the closed path
\begin{equation}
\gamma_{\downarrow} = \pi(1+\cos\theta)
\end{equation}
which is maximum for $\theta=0$,and minimum
$\gamma_{\downarrow}=0$ for $\theta=\pi$. It can be verified that
this Berry phase remains the same if we neglect the overall phase
$e^{\pm i(\phi-\chi)/2}$ from the quantized spinors as in eqns. 6
and 9 respectively. Because we obtain the identical value of Berry
phase $\gamma_{\uparrow \downarrow}$ in both the approach
identifying the same solid angle subtended about the axis of
anisotropy. It implies also that the variable $\theta$ plays the
crucial role in visualizing the Berry phase.

From the view point of quantum computation we will now proceed to
study the rotation of quantized spinor. A microscopic system such
as an atom,a nuclear spin or a polarized photon can exist in
arbitrary superposition of $\alpha|0> + \beta|1>$ where $|0>$ and
$|1>$ represent the ground and excited state respectively. In
other words any time dependent wave-function can be written as
\begin{equation}
\Psi(t)=C_0(t)|0> + C_1(t)|1>
\end{equation}
Most general pure state of a single qubit can be sufficiently
constructed using the two well known quantum gates - Hadamard
gate(H) and Phase gate as follows
\begin{equation}
~~~~~~|0>---[H]----{\bullet}^{2\theta}---[H]----{\bullet}^{\pi/2+\phi}\longrightarrow\cos\theta|0>+\sin\theta
e^{i\phi}|1>
\end{equation}

The bracketed term of the quantized spinor(equation 6), can be
written apart from the phase factor in terms of the two qubits and
the Hadamard and phase gates as follows
\begin{equation}
~~|0>---[H]---{\bullet}^{\theta}---[H]---{\bullet}^{\pi/2-\phi}\longrightarrow\cos\frac{\theta}{2}|0>+\sin\frac{\theta}{2}
e^{-i\phi}|1>
\end{equation}
Now we will focus on developing an understanding of the time
evolution of a single qubit by a general Hamiltonian. Any $2\times2$
hermitian matrix can be written in terms of unit matrix and the
three Pauli matrices.
\begin{equation}
H=\frac{\hbar}{2}(\Omega_0 \bf{1}+ \bf{\Omega}.\bf{\sigma})
\end{equation}
where $\Omega_0$ is the frequency of applied magnetic field and
$\Omega$ called the Rabi frequency which describes the transition
between the ground state $|0>$ and excited state $1>$, under the
action of the resonant field.

According to Berman, Doolen et.al [11], in the presence of rotating
magnetic field the time dependent wave-function eqn.11 are governed
by unitary transformation
\begin{equation}
\Psi(t)=U(t)\Psi(0)= U(t)[C_0(0)|0>+C_1(0)|1>]= C_0(t)|0>+C_1(t)|1>
\end{equation}
where the time dependent coefficients of qubits satisfy the
following equation.
\begin{equation}
\begin{array}{lcl}
C_0(t)&=&C_0(0)\cos\frac{\Omega t}{2}+i C_1(0)\sin\frac{\Omega t}{2}\\
C_1(t)&=&C_0(0)i\sin\frac{\Omega t}{2}+ C_1(0)\cos\frac{\Omega t}{2}
\end{array}
\end{equation}
The characteristic time of this transition $t=\pi/\Omega$ is
usually much longer than the period of precession, so that slow
and fast variables lead to some Born-Oppenheimer approximation.
This $t=\frac{\pi}{\Omega}$ is also considered as the duration of
the external field. If initially spin is in the ground state at
$(t=0)$, for $C_0(0)=1, C_1(0)=0$, the time dependent coefficients
become
\begin{equation}
C_0(t)=0,~~~~~~C_1(t)=i
\end{equation}
Thus a pulse of a resonating magnetic field with a duration
$\pi/\Omega$ drives the system from the ground state to the
excited state. Such a pulse $\pi$, conversely drives a spin from
excited to ground state also.If we apply a pulse with different
duration, we can drive the quantum system into a super-positional
state, creating a rotation of one qubit.Also when $t=\pi/2\Omega$
the resonating magnetic field has a $\pi/2$ pulse driving the
system as super-positional state of ground and excited state in
equal weight.

It seems that this rotation in the presence of magnetic field,
obviously will lead to the formation of Berry phase through the
quantization procedure. We here consider that action of $\pi$
pulse on our quantized spinor in eqn. 6 initially at the ground
state $|0>$, resulting its transfer to the excited state $|1>$,
only when
\begin{equation}
\begin{array}{lcl}
C_0(t)= \cos\frac{\theta}{2}e^{i/2(\phi-\chi)}=0\\
C_1(t)=\sin\frac{\theta}{2}e^{-i/2(\phi+\chi)}=i
\end{array}
\end{equation}
which will be possible for $$\theta=\pi ~~~and~~~~ (\phi
+\chi)/2=-\pi/2$$. This further indicate in connection with our
previous analysis in eqn.8 that only $\gamma_{\uparrow}$ will be
visible at $\theta=\pi$.

To rotate the spinor once, another $\pi$ pulse is given in order
to send the spinor from excited state back to ground one. Thus
$$C_0(0)=0 , C_1(0)=1$$ is considered initially and the action of a $\pi$
pulse to the quantized spinor result
\begin{equation}
\begin{array}{lcl}
C_0(t)= \cos\frac{\theta}{2}e^{i/2(\phi-\chi)}=i\\
C_1(t)=\sin\frac{\theta}{2}e^{-i/2(\phi+\chi)}=0
\end{array}
\end{equation}
which indicate the possibility of $$\theta=0~~~~~~ and
~~~~~~~~(\phi-\chi)/2 = \pi/2$$ visualizing  $\gamma_{\downarrow}$
only. From the above equivalence we can realize that this one
qubit rotation by the application of two $\pi$-pulses in presence
of the circularly polarized magnetic field, is equivalent to the
rotation of a quantized particle over a closed path where the
spinor changes from ground state $(|0>)$ to excited state $(|1>)$
again to ground state$(|0>)$ or the converse. Hence the net phase
acquired in this case is $(\phi+\chi+\phi-\chi)/2 =-\pi/2+\pi/2
=0$. This implies that the dynamical phase vanishes where in our
picture the variable $\phi$ is its very source. On the other hand
if $(\phi+\chi -\phi +\chi)/2= \pi/2-(-\pi/2)$, we have only the
Berry phase $\chi=-\pi$.

Thus in this course of 'spin-echo' method,the dynamical phase of
the quantized spinor can be removed if $(\phi+\chi)=-(\phi-\chi)$
is followed. For $\theta=0$, the Berry phase(BP) for spin up state
($\gamma_\uparrow$) can be removed whereas for $\theta=\pi$ the BP
for down spin ($\gamma_\downarrow$) vanishes.

\section{\bf Berry Phase in an Entangled state of two spin-1/2
particles}

 In the scheme of quantization of a Fermi field, the {\it
 direction  vector} effectively represents a vortex which is
 equivalent to a magnetic flux line. Mathematically, $\mu$ is
 associated with this magnetic flux quantum. Since our Berry phase visualizing through
$\mu$, depends on the continuous values of $\theta$, to study the
behavior of the Berry phase factor, we take the resort of the
$\mu$-theorem [12].It implies that $\mu$ can take some continuous
values where the fixed points of the Rg flow are the physical values
of the monopole strength.
 We may take $\mu$ not to be a
 fixed value but dependent on a parameter $\lambda$ and the
 function $\mu(\lambda)$ should satisfy:\\
 1.$\mu$ is stationary at the fixed points $\mu^*=\mu(\lambda^*)$ of the RG flow i,e
 $\nabla\mu(\lambda^*)=0$.\\
 2.At the fixed points $\mu(\lambda^*)$\\
 3.$\mu$ is decreasing along the infrared RG flow.i,e\\
\begin{equation}
L\frac{d\mu}{dL}\leq 0
\end{equation}
where $L$ is a length scale. We can specify
\begin{equation}
L\frac{d\mu}{dL} = -a, a\geq0
\end{equation}
Solving this we find
\begin{equation}
\mu=-alnL + c
\end{equation}
where $c$ is an arbitrary constant. Indeed neglecting the constant
term $c$, in eqn.22, we can write
\begin{equation}
\mu=-aln|x-y|
\end{equation}
when the two interacting particles are situated at the points $x$
and $y$ respectively. The effect of one on the other is depicted
through this relation in eqn 23 and its role can be observed in
the entanglement of two particles. We now will study the
appearance of Berry phase in the entanglement of two identical
spin $1/2$ quantized particles. The antisymmetric Bell state of
two spin $1/2$ particles is
\begin{equation}
|\Psi> = \frac{1}{\sqrt{2}} (|\uparrow_n>|\downarrow_n> -
|\uparrow_n>|\downarrow_n>)
\end{equation}

If we use spinor depicted by eqn.13 in the above Bell state, we
find after few mathematical steps that the entangled state
$|\Psi>$ at a particular instant $t$ is connected with the primary
entangled state $|\Psi_0>$ of qubits
\begin{equation}
|\Psi> = \cos\theta|\Psi_0> = \frac{1}{2\pi}(\gamma_{\uparrow} -
\gamma_{\downarrow})|\Psi_0>
\end{equation}
by the difference of Berry phase factor. Here $|\Psi_0>$ is
considered as the initial entangled state constructed from the
qubits $|0>$ and $|1>$.
\begin{equation}
|\Psi_0> = 1/\sqrt{2}(|1>|0> - |0>|1>)
\end{equation}

 Hence as the two quantized spin $1/2$ come close to each other,
 we have the Berry Phase of their entangled state
\begin{equation}
\gamma_{ent}= \oint <\psi|\nabla\psi>d\phi = \pi(1+\cos2\theta)
\end{equation}
 It seems that the entangled state after one rotation though acquire
Berry phase but rotation of Bell state from symmetric $\Psi_+$ to
antisymmetric $\Psi_-$ state does not takes place. The difference of
$\gamma_{\uparrow,\downarrow}$ in eqn.27 implies as if the net
topological effect due to quantization disappears. The study of
Ghrirardi et.al[13] regarding the non-entanglement of identical
fermions is different from our study due to findings after second
quantization.

With these view, two identical fermions are made quite different
before entanglement, by introducing one qubit rotation to one of
them through the circularly polarized magnetic field that plays the
role of changing the direction of quantization axis. It is done in
such a way that a resonance with the precession of the spin vector
is formed. Significantly the rotation should possess the change of
Entangled state from symmetric to antisymmetric state over the
closed path.
  This rotation of the magnetic field effectively
corresponds to the change in the direction of the magnetic flux
line. The instantaneous eigenstate of the two conjugate spin
operator in the direction of the $\vec{n(\theta/2,t)}$ expanded in
the $\sigma_z$ basis are our quantized spinors
\begin{equation}
|\uparrow(t)>=(\cos\frac{\theta}{2}|0>
+\sin\frac{\theta}{2}e^{-i\phi}|1>)e^{i(\phi-\chi)/2}
\end{equation}

\begin{equation}
|\downarrow(t)>=(\sin\frac{\theta}{2}|0>+\cos\frac{\theta}{2}e^{i\phi/2}|1>)e^{-i(\phi
+\chi)/2}
\end{equation}

For the time evolution from $t=0$ to $t=\tau$ each eigen state will
pick up a Berry phase apart from the dynamical phase [14].
\begin{equation}
|\uparrow;t=0>\rightarrow |\uparrow;t=\tau> =
e^{i\gamma_{\uparrow}(\theta)} e^{i\tilde{\theta}_+}|\uparrow_n;t=0>
\end{equation}

\begin{equation}
|\downarrow;t=0>\rightarrow |\downarrow;t=\tau> =
e^{i\gamma_{\downarrow}(\theta)}
e^{i\tilde{\theta}_-}|\downarrow_n;t=0>
\end{equation}
where $\gamma_{\uparrow,\downarrow}$ denotes the Berry phase which
is half of the solid angle swept out by the magnetic flux line and
$\tilde{\theta}_{\pm}$ is the dynamical phase. The explicit values
of these phase $\gamma_{\uparrow,\downarrow}$ are
\begin{equation}
\gamma_{\uparrow }(\theta)=\pi(1-cos\theta)=2\pi\mu_\uparrow
\end{equation}
\begin{equation}
\gamma_{\downarrow}(\theta)= \pi(1+cos\theta)=2\pi\mu_\downarrow
=2\pi-\gamma_{\uparrow}
\end{equation}
where we define
$\mu_{\uparrow,\downarrow}=\frac{1}{2}(1\mp\cos\theta)$ as the
measure of anisotropy.

If we apply spin-echo method to one spinor before entanglement with
other then the Berry phase is trapped in the entangled state,
resulting the removal of dynamical phase. This benefits to change
the entangle Bell state from antisymmetric to symmetric one at a
particular position. The most general antisymmetric Bell state for
two particles A and B situated at the points $x$ and $y$
 becomes
\begin{equation}
|\Psi(t)>=(\alpha|\uparrow(t)>|\downarrow(t)> -
\beta|\uparrow(t)>|\downarrow(t)>
\end{equation}
where $\alpha$ and $\beta$ are two complex coefficients,

With the idea of one qubit rotation of one fermion for a time
interval $\tau$ the spinor comes to its orginal state acquiring only
Berry phase and loosing the dynamical phase. The Bell state becomes
\begin{equation}
|\Psi(t=\tau)> = (e^{i\gamma_\uparrow}\alpha|\uparrow(t)>
|\downarrow(t)> - e^{i\gamma_\downarrow}\beta|\uparrow(t)>
|\downarrow(t)>
\end{equation}

 Neglecting the over all phase, we have the new form of the entangle state as
\begin{equation}
\Psi(t=\tau)=(\alpha|\uparrow(t)>|\downarrow(t)> -
e^{2i\gamma_\uparrow}|\uparrow(t)>\beta|\downarrow(t)>
\end{equation}

As we consider $\theta=\pi$ the Berry phase is removed along with
dynamical phase in the 'spin-echo' method. If we consider
$\theta=\pi/3$ then we get from the antisymmetric Bell state to
symmetric entangled state.
\begin{equation}
\Psi(t=\tau)=(\alpha|\uparrow(t)>|\downarrow(t)>
+|\beta|\uparrow(t)>|\downarrow(t)>
\end{equation}
This is analogous to the physics of fermion-boson transmutation.
Hence by varying the angle
$\theta:0\longrightarrow\pi/3\longrightarrow\pi/2$, we can
continuously change Berry phase
$\gamma:0\longrightarrow\pi/2\longrightarrow\pi$, of the
antisymmetric Bell singlet state $\Psi_-$ to the symmetric Bell
state $\Psi_+$ and back to $\Psi_-$[14].

 From the eqn. 20 we note that for very large
$L(L\longrightarrow\infty)$, $\mu$ tends to zero. Now from the
eqn. 33, we note that the limit $\mu\longrightarrow 0$ is achieved
when the angular displacement of the magnetic flux line is such
that for up spin $\cos\theta=1$ and for down spin $\cos\theta=-1$
indicating the value of $\theta=0$ to $\pi$ respectively.

Our above analysis help us to argue that Berry phase factor
${\gamma}_{\uparrow,\downarrow }=\pi(1\mp \cos\theta)$ can be
taken to be a measure of entanglement. Indeed the measure of
entanglement is usually taken to be given by the concurrence C.
 A pure general bipartite state is given by
\begin{equation}
|\psi> = \alpha|\downarrow \downarrow>+\beta|\downarrow\uparrow> +
\gamma|\uparrow\downarrow> +\delta|\uparrow\uparrow>
\end{equation}
where $\alpha,\beta,\gamma$ and $\delta$ are complex coefficients
satisfying the normalization condition. The complex concurrence is
defined by [15]
\begin{equation}
C= 2(\alpha\delta - \beta\gamma)
\end{equation}

So from eqn.(35) we note that the concurrence for the given
state(let it be up) is $C=2\beta\gamma$. Now we can relate the
norm of this concurrence with
$\mu_{\uparrow}=\frac{1}{2}(1-\cos\theta)$, the corresponding
Berry phase being $\pi(1-\cos\theta)$. Indeed when $\theta=0$ (i.e
there is no displacement of the magnetic flux line) we have
$$ |C|=0$$
which means disentanglement for up spin. For $\theta=\pi$ i.e.
there is maximum displacement of the magnetic flux line, we have
$$|C|=1$$
through the value $\mu_{\uparrow}=1$. Thus within the range of
$0\leq \mu_{\uparrow} \leq 1$ the measure of entanglement is
associated with the norm of the complex concurrence.

It is noted that when the Berry phase factor ${\mu_\uparrow}$
vanishes, we will have zero magnetic field indicating that there
is maximum disorder. Though disorder and order state depend on
temperature, when ${\mu_\uparrow}$ is maximum, we have an order
state. This suggests that we can relate the measure of
entanglement with entropy through Berry phase. Indeed there is a
relationship between the entropy after entanglement with the
concurrence which is given by
$$f(c)=H(1+\frac{\sqrt(1-C^2)}{2})$$ where $H(x)$ is the shannon
entropy. So substituting ${\mu_\uparrow}=0$ (disentanglement) we
find $f(c)=H(1)$ and for maximum entangled state
${\mu_\uparrow}=1$ yield $f(c)= H(1/2)$. Thus the maximum
entangled state for up spin can be realized through max. value of
Berry phase when the entropy is minimum indicating a highly
correlated state.

\section{\bf Noise and Berry phase}

Motivated by the works of Chiara and Palma [16] on the influence
of classical fluctuation of field on the Berry phase of spin $1/2$
particle, we now like to find the effect of noise in the Berry
phase of quantized spinor and in its entangled state both in the
presence and the absence of 'spin-echo'method. We define noise by
a shift in chirality. If we consider that with the lapse of time,
the parameter $\lambda$ suffers a deviation
$\lambda\longrightarrow \lambda+\delta\lambda$ due to any change
in $\theta$, $\phi$ and $\chi$ resulting a gauge transformation.
\begin{equation}
{\bf A(\lambda)}\longrightarrow A(\lambda) + \frac{\partial
A(\lambda)}{\partial \lambda}\delta \lambda
\end{equation}
Here $A(\lambda)$ is the gauge connection associated with the
Lagrangian in eqn.7 giving rise to Berry phase. This fluctuation
of gauge connections by the parameter $\lambda$, is the very cause
of shift in magnetic flux line corresponding chiral symmetry
breaking. Now from equation 8. considering the spin up case, we
have
\begin{equation}
A^\uparrow(\lambda)= \frac{1}{2}(1 - \cos\theta)
\end{equation}
This leads to have the noise dependent Berry connection of the
quantized spinor
\begin{equation}
{\bf A^\uparrow(\lambda)}= \frac{1}{2}(1-\cos\theta+\sin\theta
\delta \theta)
\end{equation}
which results a modification of the Berry phase
\begin{equation}
\Gamma_\uparrow=\pi(1-\cos\theta + \sin\theta
\delta\theta)=\gamma_\uparrow + \triangle\gamma
\end{equation}
and similarly for down spinor
\begin{equation}
\Gamma_{\downarrow}=\pi(1+\cos\theta -\sin\theta
\delta\theta)=\gamma_\downarrow - \triangle\gamma
\end{equation}
where we consider $\Gamma_{\uparrow,\downarrow}$ as the noise
induced Berry phase for the spin up and down quantized particles
respectively.

For the entangled state of two identical spinor, as we find in
equation 25, that the evolution of the state at a particular
instant depends on the difference of
$\gamma_{\uparrow},\gamma_{\downarrow}$ which implies increase of
noise by twice. The effect of noise in the entangled state formed
after 'spin-echo' will be less as realized from eqn.36

At the end, we like to comment that here the noise is responsible
for the fluctuation of quantization that can be applied for the
entanglement of Quantum Hall particles in the plateau and
non-plateau region.\\

{\bf Discussions}\\

We here express the quantized spinor in terms of action of quantum
gates on two qubits $|0>$ and $|1>$ that represent the ground and
excited state respectively. The one qubit rotation of
 the spinor results it to change from one state to other with the
 variation of Berry phase. The dynamical phase can be removed in the
 spin-echo method. In this method the inclusion of Berry phase in the
 entangled state is responsible for the measure of entanglement.
 Fluctuation in chirality is considered as noise that modify
 the fixed value of Berry phase.
 The effect of noise doubles as two pure identical spinor entangle.
 We like to study further this effect of noise, decoherence and
 entanglement in connection with quantization aspect of Berry phase in other quantum systems.\\

{\bf Acknowledgement}\\

DB like to acknowledge ICTP for the full assistance of this work.
DB is also grateful to Prof.Ghirardi, ICTP, and Prof. V.Vedral,
UK,and Prof  R.Resta, Universita` di Trieste for their helpful
advice. We are also thankful to Prof.B.Basu, Indian Statistical
Institute, Calcutta.

\pagebreak
%\singlespacing
\section{\bf References}
%\begin{enumerate}
\noindent [1]J.Bell, Physics {\bf 1}, 95(1964);
Rev.Mod.Phys.{\bf 38},447 (1966).\\
\noindent [2] M.A.Nielse and I.L.Chuang, {\it Quantum Computation
and Quantum Information, Cambridge University Press, Cambridge
(2000)}.\\
\noindent [3] M.V.Berry, Proc.R.Soc.London {\bf A392},45(1984). \\
\noindent [4] Y.Aharonov and J.S.Anandan, Phys.Rev.Lett, {\bf 58},
1593(1987).\\
\noindent [5]V.Vedral,quantum-ph/0212133\\
\noindent [6]V.Vedral, Cent. Eur. J. Phys. Vol 2, 289 (2003);quant-phys/0505029.\\
\noindent [7]P.bandyopadhay and K.Hajra, J. math.
Phys. Vol 28,711(1987).\\
\noindent [8] P.Bandyopadhyay; Int.J.Mod.Phys. {\bf A4}, 4449 (1989).\\
 \noindent [9] D.Banerjee and P.Bandyopadhyay; J.Math.Phys{\bf
33},990 (1992),
D.Banerjee; Fort.der Physik {\bf 44} (1996) 323\\
\noindent[10]D.Banerjee and P.Bandyopadhyay;Nuovo Cimneto,{\bf
113}(1998)921.\\
\noindent [11] G.P Berman, G.D Doolen, R. Mainnier and V.I
Tsifrinovich;{\it Introduction to Quantum Computers, World
Scientific,1998}\\
\noindent [12] P. Bandyopadhyay Int. Jor. of Mod. Phys.{\bf A15},
1415 (2000).\\
\noindent [13] G.Ghirardi, L.Marinatto and T.Weber; J. Stat.Physics {\bf 108}(2002) 49.\\
\noindent [14] R.A.Bertlmann. K.Durstberger  et.al, Phys.Rev-{\bf A69}(2004) 032112.\\
\noindent [15] W.K.Wooters; Phys.Rev.Lett {\bf 80} (1998) 2245.\\
\noindent [16] G.De Chiara and G Massino Palma ; Phys.Rev.Lett{\bf
91}, 090404 (2003).\\
%\end{enumerate}

\end{document}